\documentstyle[12pt]{article}
\normalbaselines
\begin{document}
\def\square{\hbox{\vrule\vbox{\hrule\phantom{o}\hrule}\vrule}}
\def\wh{\widehat}
\def\ov{\overline}
\renewcommand{\thefootnote}{\fnsymbol{footnote}}
\rightline{SB/F-93-204}
\begin{center}
{\Large{\bf Two gravitationally Chern-Simons \\[2mm]
terms are too many}}\\ [7mm]
{\bf C. Aragone\footnote{e-mail: aragone @usb.ve, parias
@usb.ve, adel @ciens.ula.ve

$^{\dag}$ talk presented at the 5th Canadian Conference on General Relativity
and Relativistic Astrophysics, Unv. of Waterloo, may 13-16, 1993}$^{\dag}$},
{\bf P\'{\i}o J. Arias}$^{*}$\\
{\it Departamento de F\'{\i}sica, Universidad Sim\'on Bol\'{\i}var,
Caracas}\\[4mm]
and \\ [4mm]
{\bf A. Khoudeir}$^{*}$\\
{\it Departamento de F\'{\i}sica, Universidad de los Andes, M\'erida}\\ [27mm]
{\bf Abstract}
\end{center}
It is shown that topological massive gravity augmented by the triadic
gravitational Chern-Simons first order term is a curved a pure spin-2
action. This model contains two massive spin-2 excitations. However, since its
light-front energy is not semidefinite positive, this double CS-action does not
have any physical relevance.In other words, topological massive gravity cannot
be spontaneously broken down by the presence of the triadic CS term.

\newpage

Spin needs at least two space-like dimensions. Consequently, quantum dynamical
effects for spinning particles requires space-times having dimensions not
below three. If one belives that gravity is the self-interacting massless
spin-two theory, then one has the uniform result that Einteins's action is the
appropiate model in any dimension higher than three. For $d=3$ one is faced
with the fact that there are no massless excitations having (non zero)
helicity [1]; if one believes that the essence of gravity stems in its spin
(as we do), one has to consider massive spin-2 models, as it is the case
either of topological massive gravity [2] or the recently proposed vector
Chern-Simons gravity [3]. In $d=3$ Einstein's action does not contain local
physical excitations. It cannot be taken as the unique source of a pure
spin-two theory. It is well known that however it leads to very interesting
fully topological description [4] of gravity.

In spite of this seemingly profund difference between $3-d$ Einstein's gravity
and greater than three Einstein's gravities they share a uniform property:
when a  spin-two Fierz-Pauli  mass term is added to them and goes to the
associated rigid $d$-dimensional Minkowski space time, all of them provide a
pure, (massive) spin-2 physical theory.

Since in $d=3$ the physical spin-two theory, must be massive, it is quite
straightforward to ask to all three dimensional models of gravity how they
behave of they are broken down through a Fierz-Pauli mass term. Or more in
general how they behave when some type of break-down mechanism is switched on
the initial spin-2 system.

In this talk we shall concentrate on what happens to massive spin-two
topological massive gravity (TMG) where instead of adding a Fierz-Pauli mass
term a softer break-down mechanism is present: we shall analyse what happens
to TMG wen a second Chern-Simons gravitational term breaks the initial local
Lorentz invariance of TMG.

The interest of TMG goes beyond its academic possibility. It is one of the
simplest non trivial models of a gravity theory having topological forms. It
is well known that it is a distintive feature of effective string gravities
that they contain terms having topological or quasi-topological structures like
the ten-dimensional Lorentz-Chern-Simons [5] or the Gauss-Bonet [6] type of
terms. So perhaps we might learn what are the qualitative changes arising from
the presence of topological terms in effective realistic string gravities by
analysing the novel properties (with respect to Einstein's) of $3-d$ TMG.

The action of topological massive gravity $S$ is the difference of the third
order Lorentz-Chern-Simons term $L$ and Einstein's action

$$
S = L-Et =(2\mu)^{-1}\kappa^2<\omega_p{}^a\epsilon^{prs}\partial_r
\omega_{sa}-2(3)^{-1}\epsilon^{prs}\epsilon_{abc}\omega_p{}^a \omega_r{}^b
\omega_s{}^c>
$$
$$
-\kappa^{-2}<\omega_p{}^a\epsilon^{prs}\partial_re_{sa}-2^{-1}e_p{}^a
\epsilon^{prs}\epsilon_{abc}\omega_r{}^b\omega_s{}^c>.\eqno (1)
$$
First latin letters denote Lorentz indices, middle of the alphabet world
indices, $\eta^{00}=\eta^{11}=\eta^{22}=+1$. $\omega_p{}^a,e_{sa}$ are non
independent quantities. It is understood that the affinities $\omega_p{}^a$
are determined in the standard way to be torsionless by their associated field
equations:

$$
\epsilon^{pqr}(\delta_qe_{ra}-\omega_q{}^c\epsilon_{cba}e_r{}^b)=0.\eqno (2)
$$

The Lorentz CS-term is locally conformal, Lorentz and diffeomorphism invariant
while Einstein's action is not locally conformal invariant.

$S$ can be regarded as the result of having spontaneously broken the local
conformal invariance of $L$ by means of Einstein's action. $S$ has a unique
degree of freedom having a pure massive spin-2 content.

In a curved $3-d$ space-time there are two possible CS-type of terms; the
rotational one $L$ as given in eq. (1) and the translational (or triadic)
CS-term $T$ defined in the form

$$
T =2^{-1}m\kappa^{-2}<e_p{}^a\epsilon^{prs}\partial_re_{sa}>.\eqno (3)
$$

$T$ is neither locally conformal nor local Lorentz invariant. It is just
diffeomorphism invariant. (Note that the bilinear type of CS-structure $\sim
\omega \epsilon \partial e$ appears in Einstein's action).

Let us consider the possibility of spontaneously breaking the local Lorentz
invariance of $S$ eq. (1). This can be achieved by considering the
diffeomorphism
invariant action

$$
S_{broken} = L-E+T \eqno (4)
$$

In order to analyse its content we take its quadratic part and go to Minkowski
space-time $e_p{}^a =\delta_p{}^a+\kappa h_p{}^a$, $\omega_q{}^b=\kappa
\omega_q{}^b$. $S_{broken}$ becomes

$$
S_{broken}^0= (2\mu )^{-1}<\omega_p{}^a\epsilon^{prs}\partial_r\omega_{sa}>
-2^{-1}<\omega_{pa}\omega^{ap}-\omega_a{}^a\omega_p{}^p>-
$$
$$
-2^{-1}m<h_p{}^a\epsilon^{prs}\partial_rh_{sa}>+<\lambda_p{}^a\epsilon^{prs}
(\partial_rh_{sa}-\omega_r{}^b\epsilon_{bsa})>,\eqno (5)
$$
where now $\lambda_p{}^a,\omega_p{}^a,h_p{}^a$ are independent variables.

Its equivalent third order version arises from introducing the values of
$\omega =\omega (h)$ (obtained from variations of the $\lambda$'s) into
$S_0$.

Independent variations of $\omega$, $h$, $\lambda$ yield the triplet of field
equations ($FE$).
$$
E^p{}_a =\mu^{-1}\varepsilon^{prs}\partial_r\omega_{sa}-\omega_a{}^p+
\delta_a{}^p\omega-\lambda_a{}^p+\delta_a{}^p\lambda =0,\eqno (6)
$$
$$
F^p{}_a=-m\varepsilon^{prs}\partial_rh_{sa}+\varepsilon^{prs}\partial_r
\lambda_s{}^a=0,\eqno (7)
$$
and
$$
G^p{}_a =\varepsilon^{prs}\partial_rh_{sa}-\omega_a{}^p+\delta_a{}^p
\omega =0.\eqno (8)
$$

Considering the lower spin sector of these eqs., i.e. computing
$E \equiv E^p{}_p$, $F$, $G$, $\partial_pE^p{}_a,\cdots$ and
$\varepsilon_{pab}E^{pa}\equiv  \check{E}_b$, $\check{F}_b$, $\check{C}_b$
it is straightforward to see that this system only propagates spin-2
excitations.

Both, the spin-1
$\varepsilon_{pab}\omega^{pa},\cdots,\varepsilon_{pab}\lambda^{pa},
\partial_p\omega_{pa},\partial_p\lambda_{pa}$
and the scalar sector of $\omega ,h, \lambda$ vanish in the harmonic gauge
$\partial_ph_{pa}=0$.

Projection of the $FE$ (7) (8) (9) upon the spin-2$^+$ (spin-2$^-$) subspaces
using the pseudospin-2$^\pm$ projectors [7], gives
$$
(X-1)\omega^{T+}-\lambda^{T+}=0\  \ ,\  \ -mXh^{T+}+X\lambda^{T+}=0\  \ , \  \
\omega^{T+}=Xh^{T+} \eqno (9)
$$
where $X=\mu^{-1}\square^{1/2},\mu =1$, $m$ means the dimensionless relation
$m\mu^{-1}$ and $h^{T+}$ denotes the spin-2$^+$ part of $h_{pa}$.

The inverse propagator is therefore
$$
\Delta^+(X)=X[X(X-1)-m].\eqno (10)
$$

There is a positive mass $m=2^{-1}+(4^{-1}+m)^{1/2}$ in the spin-2$^+$ sector.
Similarly, since $\Delta^-(X)=X[X(X+1)-m]$ we might have a spin-2$^-$
excitation with mass $m^-=-2^{-1}+(4^{-1}+m)^{1/2}$.

We want to see whether this system has its energy bounded from below (or not).
It will be shown that, independently of the sign of $m$, the light-front (LF)
generator is unbounded and consequently action (6) is physically meaningless,
in spite of the fact that, from a covariant point of view, the system (7),
(8), (9), seems to propagate two spin-2 decoupled excitations.

In order to have this, we calculate the value of the LF-generator of action
(6) in terms of its two unconstrained variables $\omega_{vv}$ and
$\lambda_{vv}$. Light front coordinates $(u,v)$ are defined by
$$
\eta^{11}=1=-\eta^{uv},\  \ u =2^{-1/2}(x^0-x^2),\  \
v =2^{-1/2}(x^0+x^2),\  \ \varepsilon^{1vu}=+1.\eqno (11)
$$
Time derivatives are written $\partial_u f=\dot{f}$ and the LF-spacelike ones
are denoted $\partial_vf=f'$

One starts from the covariant expressions (6) of $S_0$ and express this action
in terms  of the 27 $LF$-field components $\omega_{uu}\equiv \omega_u$,
$\omega_{uv}$, $\omega_{vu}$, $\omega_v \equiv\omega_{vv}$,
$\omega_1 \equiv\omega_{11}$, $\omega_{1u}$, $\omega_{u1}$, $\omega_{1v}$ and
$\omega_{v1},\cdots ,\lambda_u,\lambda_{uv},\cdots ,\lambda_{1v},\lambda_{v1}$.

It is inmediate to realize that $\omega_{ua},h_{ub},\lambda_{uc}$ are
multipliers associatted with nine differential constraint equations which
can be solved, providing the values of $\omega_{1a},h_{1b},\lambda_{1c}$ as
functions of the remaining nine intermediate variables
$\omega_{va},h_{vb},\lambda_{vc}$. Their solution is:
$$
\wh{\omega}_{1v}=(\partial_1+1)\wh{\omega}_v+\wh{\lambda}_v
\  \ ,\  \ \wh{h}_{1v}=\partial_1\wh{h}_v+\wh{\omega}_v,\eqno (12a,b)
$$
$$
\wh{\lambda}_{1v}=\partial_1\wh{\lambda}_v+m\wh{\omega}_v,
\eqno (12c)
$$
$$
\omega_1=\partial_1\wh{\omega}_{v1}+\wh{\omega}_{1v}+
\wh{\lambda}_{1v}\  \ ,\  \ h_1=\partial_1\wh{h}_{v1}+
\wh{\omega}_{1v},\eqno (13a,b)
$$
$$
\lambda_1=\partial_1\wh{\lambda}_{v1}+m\wh{\omega}_{1v},\eqno (13c)
$$
$$
\omega_{1u}{}'=(\partial_1-1)\omega_{vu}-\lambda_{vu}+\partial_1
\wh{\omega}_{v1}+\partial_1\wh{\lambda}_{v1}+(m+1)
\wh{\omega}_{1v}+\wh{\lambda}_{1v}\eqno (14a)
$$
$$
h_{1u}{}'=\partial_1h_{vu}-\omega_{vu}+\partial_1\wh{\omega}_{v1}+
\wh{\omega}_{1v}+\wh{\lambda}_{1v}\eqno (14b)
$$
$$
\lambda_{1u}{}'=\partial_1\lambda_{vu}-m\omega_{vu}+
m\partial_1\wh{\omega}_{v1}+m\wh{\omega}_{1v}+
m\wh{\lambda}_{1v}\eqno (14c)
$$
where we introduced redefinitions like $\omega_{v1}\equiv
\wh{\omega}'_{v1}$,  $\omega_{v}\equiv \wh{\omega}''_{v}$,
$\omega_{1v}\equiv \wh{\omega}'_{1v}$ for the
three sets of variables $\omega$, $\lambda$, $h$.

In principle, the intermediate expression of $S_0$ obtained in terms of the
nine intermediate variables $\omega_{va},h_{vb},\lambda_{vc}$ might have
$\omega_{vu},h_{vu},\lambda_{vu}$ in the dynamical germ (the piece of
$S_0\sim p\dot{q}$).

However it turns out after using eqs. (13), (14) that these three variables
are not present in this part of the action. While $\omega_{vu}$ and
$\lambda_{vu}$ constitute two additional Lagrange multipliers, $h_{vu}$ has
totally disappeared.

Independent variations of $\omega_{vu},\lambda_{vu}$ lead to the final two
differential constraints of $S_0$. Their solution shows the symmetry of the
$1v$-components $\omega_{1v},\lambda_{1v}$, i.e.
$$
\wh{\omega}_{v1}=\wh{\omega}_{1v}\  \ ,\  \
\wh{\lambda}_{v1}=\wh{\lambda}_{1v}\eqno (15a,b)
$$

Now it is immediate to obtain the unconstrained form of the evolution
generator $G$ of action $S_0\sim p\dot{q}-G$. Since, at the initial stage
when one writes down $S_0$ in terms of the LF-variables, $G$ had the form:
$$
G=<(\wh{\omega}_{v1}+\wh{\lambda}_{v1})\omega_{1v}{}'+\wh{\omega}_{v1}
\lambda_{1u}{}'>;\eqno (16)
$$
it is straightforward to realize that, after insertion of the values (15) of
$\omega_{1u}{}',\lambda_{1u}{}'$ in it, $G$ becomes:
$$
G=<[(m+1)\wh{\omega}_{1v}+\wh{\lambda}_{1v}]^2-m^2\wh{\omega}^2_{1v}>.
\eqno (17)
$$
This explicitly shows that the generator is a non semidefinite positive
quadratic expression. Consequently the unconstrained reduced form of
$S_0$, even written in terms of the unique two gauge-invariant variables
$\wh{\lambda}_v,\wh{\omega}_v$, does not have physical relevance.
One can say that the presence of both types of $CS$ terms is inconsistent. This
situation is peculiar of Lorentz-$CS$ gravity (there is no analogous third
order  $CS$ theory for vector fields).

As we have a covariant first order action for studying this system, one can
also perform a canonical newtonian $2+1$ type of analysis which leads to the
analogous physical result; the energy is not definite positive. One might also
wonder whether the  pure double CS-gravitational action

$$
S' =L+\varepsilon T \eqno (18)
$$
has some physical relevance. Its quadratic approximation on flat $3d$ Minkowski
space consists of

$$
S'{}^Q = (2\mu )^{-1}<\omega_p{}^a\epsilon^{prs}\partial_r\omega_{sa}> +
\varepsilon 2^{-1}m<h_p{}^a\epsilon^{prs}\partial_rh_{sa}>+
$$
$$
+<\lambda_p{}^a\epsilon^{prs}(\partial_rh_{sa}-\omega_r{}^b\epsilon_{bsa})>.\eqno
(19)
$$

Doing a similar analysis one finds out that the system only contains spin-2
excitations. However its energy is unbounded from below and consequently
$S'=L+\varepsilon T$ has no physical relevance.

\newpage

\begin{center}
{\bf References}
\end{center}
\begin{enumerate}
\item B.Binegar, J. Math. Phys.{\bf 23} (1982) 1511

\item S. Deser, R. Jackiw and S. Templeton, Ann. of Phys.(N.Y) {\bf 140} (1982)
372.

\item C. Aragone, P. J. Arias and A. Khoudeir Preprint SB/F-92-192 {\it Massive
Vector Chern-Simons gravity}.

\item S. Deser, t'Hooft, R. Jackiw Ann. of Phys.(N.Y) {\bf 152} (1984) 220

\item A. Chamseddine Nucl. Phys. {\bf B185} (1981) 403.

E. Bergshoeff, M. de Roo, B. de Wit and P. van Nieuwenhuizen Nucl. Phys {\bf
B195} (1982) 97.

G. F. Chapline ans N. S. Manton Phys. Lett. {\bf B120} (1983) 105.

\item B. Zumino, Phys. Rep. {\bf 137} (1986) 109.

\item C.Aragone and A.Khoudeir Phys. Lett.{\bf B173 }(1986) 141

\end{enumerate}

\end{document}